# THE OVERVIEW OF THE NATIONAL IGNITION FACILITY DISTRIBUTED COMPUTER CONTROL SYSTEM

L. J. Lagin, R. C. Bettenhausen, R. A. Carey, C. M. Estes, J. M. Fisher, J. E. Krammen, R. K. Reed, P. J. VanArsdall, J. P. Woodruff, LLNL, Livermore, CA 94551, USA

Abstract

The Integrated Computer Control System (ICCS) for the National Ignition Facility (NIF) is a layered architecture of 300 front-end processors (FEP) coordinated by supervisor subsystems including automatic beam alignment and wavefront control, laser and target diagnostics, pulse power, and shot control timed to 30 ps. FEP computers incorporate either VxWorks on PowerPC or Solaris on UltraSPARC processors that interface to over 45,000 control points attached to VME-bus or PCI-bus crates respectively. Typical devices are stepping motors, transient digitizers, calorimeters, and photodiodes. The front-end layer is divided into another segment comprised of an additional 14,000 control points for industrial controls including vacuum, argon, synthetic air, and safety interlocks implemented with Allen-Bradley programmable logic controllers (PLCs). The computer network is augmented asynchronous transfer mode (ATM) that delivers video streams from 500 sensor cameras monitoring the 192 laser beams to operator workstations. Software is based on an object-oriented framework using CORBA distribution that incorporates services for archiving, machine configuration, graphical user interface, monitoring, event logging, scripting, alert management, and access control. Software coding using a mixed language environment of Ada95 and Java is one-third complete at over 300 thousand source lines. Control system installation is currently under way for the first 8 beams, with project completion scheduled for 2008.

## 1 INTRODUCTION

This paper presents an overview of the NIF ICCS. The NIF contains 192 laser beam lines that are focused on an inertial confinement fusion (ICF) capsule at target chamber center [1]. Each beam requires alignment, diagnostics, and control of power conditioning and electro-optic subsystems. NIF will be capable of firing target shots every 8 hours, allowing time for the components to cool sufficiently to permit precise realignment of the laser beams onto the target.

The NIF requires integration of about 60,000 atypical control points, must be highly automated and robust, and will operate around the clock. Furthermore, facilities such as the NIF represent major capital investments that will be operated, maintained, and upgraded for decades. Therefore, the computers and control subsystems must be relatively easy to extend or replace periodically with newer technology.

The ICCS architecture was devised to address the general problem of providing distributed control for large scientific facilities that do not require real-time capability within the supervisory software. The ICCS architecture uses the client–server software model with event-driven communications. Some real-time control is also necessary; controls requiring deterministic response are implemented at the edges of the architecture in front-end computer equipment. The software architecture is sufficiently abstract to accommodate diverse hardware, and it allows the construction of all the applications from an object-oriented software framework that will be extensible and maintainable throughout the project life cycle. This framework offers interoperability among different computers and operating systems by leveraging a common object request broker architecture (CORBA). The ICCS software framework is the key to managing system complexity.

A brief summary of performance and functional requirements follows in Table 1.

Table 1: Selected ICCS performance requirements

| Computer restart | < 30 minutes |
|---|---|
| Post-shot data recovery | < 5 minutes |
| Respond to broad-view status updates | < 10 seconds |
| Respond to alerts | < 1 second |
| Perform automatic alignment | < 1 hour |
| Transfer and display digital motion video | 10 frames per second |
| Human-in-the-loop controls response | within 100 ms |

Summary ICCS functional requirements:
- Provide graphical operator controls and equipment status.
- Maintain records of system performance and operational history.
- Automate predetermined control sequences (e.g., alignment).
- Coordinate shot setup, countdown, and shot data archiving.
- Incorporate safety and equipment protection interlocks.

# 2 ARCHITECTURE

The ICCS is a layered architecture consisting of FEPs coordinated by a supervisory system (Figure 1). Supervisory controls, which are hosted on UNIX workstations, provide centralized operator controls and status, data archiving, and integration services. FEP computers incorporate either VxWorks on PowerPC or Solaris on UltraSPARC processors that interface to over 45,000 control points attached to VME-bus or PCI-bus crates respectively. Typical devices are stepping motors, transient digitizers, calorimeters, and photodiodes. FEP software provides the distributed services needed to operate the control points by the supervisory system. Functions requiring real-time implementation are allocated to software within the FEP (or embedded controllers) that does not require time-critical communication over the local area network. Precise triggering of 1600 channels of fast diagnostics and controls is handled during a 2-second shot interval by the timing system, which is capable of providing triggers to 30-ps accuracy and stability [2]. The software is distributed among the computers and provides plug-in software extensibility for attaching control points and other software services by using CORBA.

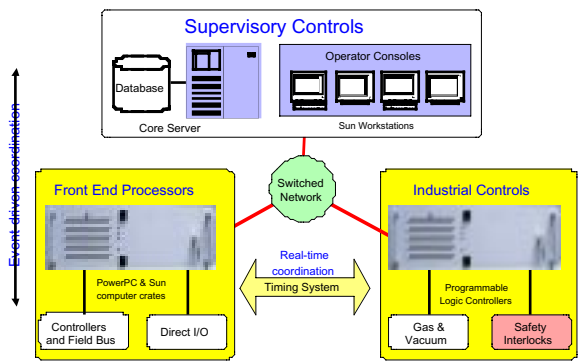

**Figure 1: Control system architecture.**

Operator consoles provide the human interface in the form of operator displays, data retrieval and processing, and coordination of control functions. Supervisory software is partitioned into several cohesive subsystems, each of which controls a primary NIF subsystem such as alignment or power conditioning. Several databases and file servers are incorporated to manage both experimental data and data used during operations and maintenance.

FEPs implement the distributed portion of the ICCS by interfacing to the NIF control points. The FEP software performs sequencing, data acquisition and reduction, instrumentation control, and input/output operations. The software framework includes a standard way for FEPs to be integrated into the supervisory system by providing the common distribution mechanism coupled with software patterns for hardware configuration, status, and control.

The front-end layer is also divided into another segment comprised of an additional 14,000 control points for industrial controls including vacuum, argon, synthetic air, and safety interlocks implemented with Allen-Bradley PLCs. The segment consists of a network of PLCs that reside below the FEP layer and attach to field devices controlling vacuum systems for the target chamber and spatial filters, argon gas controls for the beam tubes, and thermal gas conditioning for amplifier cooling. This segment also monitors the independent safety interlock system, which monitors doors, hatches, shutters, and other sensors in the facility.

ICCS is further divided into subsystems to partition activity and ensure performance. There are ten supervisor software applications that conduct NIF shots in collaboration with 17 kinds of FEPs. Software applications in the NIF control system subsystems partition the control system into loosely coupled vertical slices consisting of a supervisor and associated FEPs that are easier to design, construct, operate, and maintain. All supervisors are controlled by a Shot Director, which is responsible for conducting the shot plan, distributing the countdown clock, and coordinating the other subsystems. The beam control supervisor provides coordination and supervision of laser wavefront control and laser component manual and automatic alignment and optics inspection [3]. The laser diagnostics supervisor provides functions for diagnosing laser performance by collecting integrated, transient and image information from sensors positioned in the beams [4]. The optical pulse generation supervisor provides temporally and spatially formatted optical pulses with the correct energetics and optical characteristics required for each of the beams. The target diagnostics supervisor coordinates the collection of data from a diverse and changing set of instruments. The power conditioning supervisor manages high-voltage power supplies that fire the main laser amplifiers. The Pockels cell supervisor manages operation of the plasma electrode Pockels cell optical switch that facilitates multipass amplification within the main laser amplifiers. The shot services supervisor provides monitoring of industrial controls and integrated timing systems. A final supervisor interfaces to a computerized Laser Performance Operations Model (LPOM) simulation, which is being developed to guide laser setup of laser operating parameters [5].

## 3 COMPUTER SYSTEM AND NETWORK

A computer network will interconnect approximately 750 systems, including embedded controllers, 300 FEPs, supervisory workstation systems, and centralized servers [6]. All systems are networked with Ethernet to serve the majority of communication needs, and ATM is utilized to transport multicast video and synchronization triggers. CORBA software infrastructure provides location-independent communication services over TCP/IP between the application processes in the workstation supervisors, servers, and FEPs. Video images sampled at a 10-Hz frame rate from any of 500 video cameras will be multicast using direct ATM data streams from video FEPs to any operator console.

## 4 ICCS SOFTWARE FRAMEWORK

The ICCS is based on a scalable software framework that is distributed over supervisory and FEP computers throughout the NIF facility [7]. The framework provides templates and services at multiple levels of abstraction for the construction of software applications that distribute via CORBA. Framework services such as alerts, events, message logging, reservations, user interface consistency, and status propagation are implemented as templates that are extended for each by application software. Application software is constructed on a set of framework components that assure uniform behavior spanning the FEP and supervisor programs [8]. Additional framework services are provided by centralized server programs that implement database archiving, name services, and process management.

## 5 SOFTWARE DEVELOPMENT AND TESTING

The ICCS incorporates a mixed language environment of Ada95 and Java using CORBA and object-oriented techniques to enhance the openness of the architecture and portability of the software. The object-oriented design is captured in Unified Modeling Language using the Rational Rose design tool that maintains schematic drawings of the software architecture, while source code is managed by the Rational Apex tool.

The strategy used to develop the NIF ICCS calls for incremental cycles of construction and formal test to deliver an estimated total of 1 million lines of code [9]. Each incremental release allocates four to six months to implement targeted functionality and culminates when offline tests are conducted in the ICCS Integration and Test Facility. Tests are then repeated online to confirm integrated operation in dedicated laser laboratories or ultimately in the NIF. Process measurements including earned-value, product size, and defect densities provide software project controls and process improvements that generate confidence that the control system will be successfully deployed.

## 6 SUMMARY

Construction of the NIF ICCS incorporates many of the latest advances in distributed computer and object-oriented software technology. Primary goals of the design are to provide an open, extensible, and reliable architecture that can be maintained and upgraded for decades. The control system is being developed using the incremental approach to software construction. Eighteen (of over 40 planned) releases of software have been deployed and tested since 1998, resulting in over 325 thousand lines of source code that is capable of supporting initial laser operations. Control system installation is under way for the first 8 beams, with project completion scheduled for 2008.